\RequirePackage{lineno}
\documentclass[aps,prl,twocolumn,superscriptaddress,groupedaddress,preprintnumbers]{revtex4}  
\usepackage{graphicx}  
\usepackage{dcolumn}   
\usepackage{bm}        
\usepackage{amssymb}   
\usepackage{multirow}
\usepackage{epsfig}
\usepackage{amsmath}
\usepackage[colorlinks, linkcolor=blue]{hyperref}
\usepackage{ulem}

\newcommand{\effstw}{\ensuremath{\sin^2\theta_{\text{eff}}^{\text{$\ell$}}}}

\newcommand{\afb}{A_\text{FB}}

\begin{document}

\lefthyphenmin=2
\righthyphenmin=2

\widetext

\title{Relative difference between up and down quark structure of the proton }
\affiliation{Department of Modern Physics, University of Science and Technology of China, Jinzhai Road 96, Hefei, Anhui 230026, China}
\affiliation{Department of Physics and Astronomy, Michigan State University, East Lansing, MI 48823, USA}

\author{Mingzhe Xie} \affiliation{Department of Modern Physics, University of Science and Technology of China, Jinzhai Road 96, Hefei, Anhui 230026, China}
\author{Siqi Yang}\email{yangsq@ustc.edu.cn} 
\affiliation{Department of Modern Physics, University of Science and Technology of China, Jinzhai Road 96, Hefei, Anhui 230026, China}
\author{Wenhao Ma}\affiliation{Department of Modern Physics, University of Science and Technology of China, Jinzhai Road 96, Hefei, Anhui 230026, China}
\author{Minghui Liu} \affiliation{Department of Modern Physics, University of Science and Technology of China, Jinzhai Road 96, Hefei, Anhui 230026, China}
\author{Liang Han} \affiliation{Department of Modern Physics, University of Science and Technology of China, Jinzhai Road 96, Hefei, Anhui 230026, China}
\author{C.-P. Yuan} \affiliation{Department of Physics and Astronomy, Michigan State University, East Lansing, MI 48823, USA}

\begin{abstract}
We present a novel determination of the down-to-up quark composition ratio using the forward-backward asymmetry 
observed in the proton-proton collisions at the LHC. This method offers unique insights into the flavor-specific 
difference between down and up quarks, which are difficult to isolate  in traditional cross-section measurements 
due to the inherent mixing of contributions from both flavors. In this study, we systematically measure the down-to-up 
quark ratio over a broad momentum fraction ($x$) range of 0.01 to 0.1, utilizing the sensitivity of the forward-backward 
asymmetry to quark-level couplings. Our findings reveal significant deviations in both the value and $x$-dependence 
of this ratio compared to predictions from current parton distribution functions (PDFs). These discrepancies 
highlight potential limitations in existing PDF parameterization and emphasize the importance of flavor-separated 
measurements for advancing our understanding of proton structure.

\end{abstract}
\maketitle

The relative difference between the up and down quark distributions 
is of importance in both the proton structure and quantum 
chromodynamic (QCD) studies. 
For the ``valence'' component at large momentum fraction of $x\sim 0.1$, 
$u$ and $d$ quark densities have significant non-perturbative contributions, 
of which the ratio is related to the constituent 
description of the proton. For the ``sea'' component,   
the relative difference between 
$u$($\bar{u}$) and $d$($\bar{d}$) directly reflects 
the impact of both the non-perturbative and perturbative (arising from gluon 
splitting) QCD dynamics.  
Therefore, the relative difference 
between $u(\bar{u})$ and $d(\bar{d})$ parton distribution functions 
(PDFs), as a function of the parton momentum fraction $x$ and the 
energy scale $Q$, is essential to the studies of both proton structure and 
perturbative QCD. 
Although various measurements 
have been performed to study 
the up and down quark distributions in past decades, direct observations 
of the relative difference in these PDFs are very rare. 
It is because most physical interactions, 
from the lepton-nucleon/neutrino-nucleon 
deep inelastic scattering (DIS) to the hadron-hadron Drell-Yan process,  
can happen via both 
flavors with comparable cross sections and similar final state particles, 
so that the
$u(\bar{u})$ and $d(\bar{d})$ contributions always mix  
and are indistinguishable. 
For example, in the latest global 
analysis of PDFs~\cite{CT18, MSHT20, NNPDF4}, 
such mixture dominates in 
the measurements of both the $Z$ boson productions 
~\cite{HERA, D0Z, CDFZ, LHCb7TeVZ, LHCb8TeVZ, ATLAS7TeVZ, ATLAS8TeVZ} 
and the $\gamma^*$-exchanged interactions
~\cite{HERA, BCDMSproton, BCDMSdeut, NewMuon, diMuon-pc, NuSeaRatio, NuSeaXsection}. 
Even in the charged-current $W^{\pm}$ boson productions, the observed 
cross sections can still be a mixture of $u$ and $\bar{d}$, or $d$ and $\bar{u}$
~\cite{CDFW, D0Wee, CMS7TeVW, ATLAS7TeVW, D0Wmumu, CMS8TeVW}. 

In this Letter, we report a determination of the relative difference between the 
up and down quark contributions from the forward-backward asymmetry ($\afb$) 
of the $pp\rightarrow Z/\gamma^* \rightarrow \ell^+\ell^-$ process. 
It has been shown in Ref.~\cite{AFBFactorization} that 
the $u$($\bar{u}$) and $d$($\bar{d}$) quark information in $\afb$ can be 
separately factorized into new structure parameters, $P_u$ and $P_d$, without 
any mixture of the two types. The ratio, $R=P_d/P_u$, 
thus reflects the unique information of the relative difference between up and down 
quarks, of which the contributions of $s$, $c$ and $b$ quarks nearly perfectly cancelled. 
This is because $q(x,Q)=\bar{q}(x,Q)$, for $q=s$, $c$, and $b$, up to the next-to-leading 
order (NLO) in QCD interactions, assuming that $q(x, Q_0)=\bar{q}(x, Q_0)$ 
at the initial scale $Q_0$ around 1 GeV. 
This new method has already been applied at both the LHC and Tevatron. In 
the first application of 
Ref.~\cite{CMS8TeVExt}, $P_u$ and $P_d$ were extracted from the $\afb$ distributions 
reported by the CMS collaboration with their 8 TeV data~\cite{CMSAFB8TeV}, giving a 
tendency of $R$ to be higher than the current PDF predictions around $x>0.05$. 
However, the precision in $R$ 
was limited because the effective weak mixing angle $\effstw$ was simultaneously fitted as an 
additional free parameter, which smeared the sensitivity to the parton information. 
Following that, the D0 collaboration measured $P_u$ and $P_d$ using the data collected 
during its Run IIb period~\cite{D0PuPd}, giving $R$ higher than the PDF predictions by 
more than 3.5 standard deviations at $x\sim 0.1$. In the D0 measurement, $\effstw$ was fixed 
to its experimentally determined value and the corresponding uncertainty 
arising from varying its value to the determination of $R$ 
 is found to be negligible. 
 
In this work, we measure 
$P_u$, $P_d$ and $R$ using the latest $\afb$ measurement reported by the 
CMS collaboration with 138 fb$^{-1}$ data at 13 TeV~\cite{CMS13TeVAFB, CMS13TeVAFBMetaData}. 
We also re-measure the $R$ parameter using the previous CMS 8 TeV $\afb$ with 
$\effstw$ value fixed to improve the precision of the measurement. 
As to be illustrated later, this new determination gives precise observation in a 
wide range of $x$. The $R$ parameter is found 
to be significantly higher than the current PDF predictions when $x>0.05$, while 
at $x\approx 0.01$ it is instead lower than the predictions. 
The new results not only give large deviation in the down-to-up ratio, but also 
indicate a different shape in the $x$ dependence. 
 The deviation also reflects 
the fact that direct constraint on the up-to-down relative difference is very weak 
in current PDF global analysis.
As a new observation with unique information, the findings from this work 
is expected to have important impact on both the proton structure study and the perturbative 
QCD predictions.
 
According to Ref.~\cite{AFBFactorization}, $\afb$ in the 
$pp\rightarrow Z/\gamma^*\rightarrow \ell^+\ell^-$ process can be factorized up to all orders as:

\begin{footnotesize}
\begin{eqnarray}\label{eq:LHCAFBfactorization1}
\afb(Y, M, Q_T) &=& C_u(Y, M, Q_T)A^u_\text{FB}(Y, M, Q_T; \effstw) \nonumber\\
   &+ & C_d(Y, M, Q_T)A^d_\text{FB}(Y,M,Q_T;\effstw), 
\end{eqnarray}
\end{footnotesize}
 
\noindent where $Y$, $M$ and $Q_T$ are the rapidity, invariant mass, and 
transverse momentum of the di-lepton system. $C_u$ and $C_d$ are the 
structure parameters representing the parton information in the initial state. 
$A^u_\text{FB}$ and $A^d_\text{FB}$ 
are the forward-backward asymmetries corresponding to the  
$u\bar{u}\rightarrow Z/\gamma^*\rightarrow \ell^+\ell^-$ and 
$d\bar{d}\rightarrow Z/\gamma^*\rightarrow \ell^+\ell^-$ hard processes. They are independent 
of the parton information in the initial state, and can be precisely predicted. 
 $A^u_\text{FB}$ and $A^d_\text{FB}$ are naturally different, especially 
as functions of the invariant mass of the lepton pair $M$ 
around the $Z$ boson mass pole. Therefore, $C_u$ and $C_d$ 
can be simultaneously determined from the observed $\afb$ distribution.

In each $Y$-$M$-$Q_T$ interval,  
the structure parameters can be written in terms  
of parton densities as:

\begin{footnotesize}
\begin{eqnarray}\label{eq:LHCAFBfactorization2}
 C_u &=&  \frac{ \left[ u(x_1)\bar{u}(x_2)-u(x_2)\bar{u}(x_1) \right] \mathcal{N}_u}{\sum_{q=u,c,d,s,b} \left[ q(x_1)\bar{q}(x_2)+q(x_2)\bar{q}(x_1) \right] \mathcal{N}_q }\nonumber \\
 C_d &=&  \frac{ \left[ d(x_1)\bar{d}(x_2)-d(x_2)\bar{d}(x_1) \right] \mathcal{N}_d}{\sum_{q=u,c,d,s,b} \left[ q(x_1)\bar{q}(x_2)+q(x_2)\bar{q}(x_1) \right] \mathcal{N}_q }.
\end{eqnarray}
\end{footnotesize}

\noindent where $\mathcal{N}_q$ are 
the event rates contributed by the hard scattering processes involving 
$q-\bar{q}-Z/\gamma^*$ couplings, and $C_u$ and $C_d$ contain all 
higher order QCD corrections included in the factorized form of Eq.~\eqref{eq:LHCAFBfactorization1}, 
with the parton momentum fractions $x_1$ and $x_2$ defined by

\begin{footnotesize}
\begin{eqnarray}\label{eq:LOx}
  x_{1,2} = \frac{\sqrt{M^2 + Q^2_T}}{\sqrt{s}} \times e^{\pm Y}
\end{eqnarray}
\end{footnotesize}

\noindent where $\sqrt{s}$ is the collider energy. Therefore, it is equivalent to 
have $(x_1,x_2)$ and $(Y,M,Q_T)$ to describe the dependences. 
$C_u$ and $C_d$ contain
only the up and  down quark information in their numerators. 
The denominator is  
the total cross section of the Drell-Yan process, which cancels in 
the ratio of $R = C_d/C_u$. 
Therefore, $R$ is an observable purely related to the relative difference between 
the $d(\bar{d})$ and $u(\bar{u})$ contributions. 
In the LHC's high energy collisions, $x_2$ is usually 
at $\mathcal{O}(10^{-3})$ where  
the perturbative QCD contribution becomes dominant in the light quark 
densities, resulting in consistent distributions of light quarks within their 
uncertainties. 
At $x_1>0.01$ however, the non-perturbative contributions still can cause 
sizable difference between light quarks~\cite{CT18}. 
Taking the assumption of 
$u(x_2)\approx \bar{u}(x_2) \approx d(x_2) \approx \bar{d}(x_2)$, 
we have 
$R \approx (d(x_1) - \bar{d}(x_1)) / (u(x_1) - \bar{u}(x_1)) = d_V(x_1)/u_V(x_1)$, 
which is an experimental observable nearly perfectly reflecting the valence quark ratio 
without any mixture of other quark information.  

Given the fact that the $(Y,M,Q_T)$ intervals cannot be infinitely small in practice, 
it is the average structure 
parameters, $P_u$ and $P_d$, actually being measured:

\begin{footnotesize}
\begin{eqnarray}\label{eq:PuPd}
  C_u(Y, M, Q_T) &=& P_u + \Delta_u(Y, M, Q_T) \nonumber \\
  C_d(Y, M, Q_T) &=& P_d + \Delta_d(Y, M, Q_T).
\end{eqnarray}
\end{footnotesize}

\noindent $\Delta_u(Y, M, Q_T)$ and $\Delta_d(Y, M, Q_T)$, which describe the 
kinematic dependences 
in that interval,  
are usually fixed, and defined to yield zero when integrated over $Y$, $M$, and $Q_T$. 
In principle, fixing $\Delta_u$ and $\Delta_d$ introduces uncertainties on $P_u$ and $P_d$.  
However, as demonstrated in Ref.~\cite{AFBFactorization, CMS8TeVExt, D0PuPd}, such uncertainties can be 
small with current binning of experimental observations.
 
In the recent CMS 13 TeV new measurement, the 
asymmetry of the Drell-Yan events is  
measured as the effective angular coefficient $A_4$, defined in the cross section:
 
\begin{footnotesize}
\begin{eqnarray}\label{eq:DrellYanXsection}
\frac{\text{d}\sigma}{\text{d}\cos\theta \text{d}Y \text{d}M} &\propto & 1+\cos\theta^* + A_4(Y, M) \cos\theta^* \nonumber \\
   & &  + \frac{1}{2} A_0(Y, M) (1-3\cos^2\theta^*)
\end{eqnarray}
\end{footnotesize}
 
\noindent in full phase space. $\theta^*$ is the scattering angle of the event defined 
in the Collins-Soper frame~\cite{CSframe}. By this definition, 
the overall $A_4$ directly relates to the forward-backward 
asymmetry as $\afb = (3/8) A_4$. For simplicity, we do not distinguish $\afb$ and $A_4$  
in the rest of this paper as they differ by only a constant factor. 
$\afb$ is observed in 9 intervals of $|Y|$ from 0 to 3.4, and in a mass window 
not wider than 100 GeV around the $Z$ boson mass pole. Although $\afb$ are observed 
in different mass bins, $P_u$ and $P_d$ can only be 
measured in function of $|Y|$ accordingly, because the simultaneous fit on $P_u$ and $P_d$ relies 
on the information of the difference in the mass dependence between $A^u_\text{FB}$ and $A^d_\text{FB}$~\cite{AFBFactorization}.
Therefore, in this study we measure $P_u$ and $P_d$ as averages over the $Z$ boson mass pole in 
each $|Y|$ interval. 

The central values of $P_u$ and $P_d$ in each $|Y|$ interval are determined by requiring the 
best agreement between the experimentally measured $\afb$ and the predicted $\afb$ in Eq.~\eqref{eq:LHCAFBfactorization1} 
and Eq.~\eqref{eq:PuPd} via the minimum $\chi^2$ fit:

\begin{footnotesize}
\begin{eqnarray}
 \chi^2 = \sum_i \frac{\left(A^\text{data}_\text{FB}(M_i) - A^\text{pred.}_\text{FB}(M_i)\right)^2}{\sigma^2(M_i)}
\end{eqnarray}
\end{footnotesize}

\noindent According to the CMS report, the bin-by-bin correlations of the total uncertainties $\sigma(M_i)$ are 
negligible~\cite{CMS13TeVAFBMetaData}.
In the calculations of $A^\text{pred.}_\text{FB}$, 
the hard-process-level asymmetries $A^u_\text{FB}$ and 
$A^d_\text{FB}$ are computed using the 
{\sc ResBos}~\cite{resbos} package in which the QCD interaction is calculated at 
approximate next-to-next-to-leading order 
(NNLO) plus next-to-next-to-next-to-leading logarithm (N$^3$LL), and the electroweak (EW) 
interaction is 
calculated based on the effective born approximation~\cite{PDG}, 
which gives precise predictions on the relationship between $\afb$ and $\effstw$ around the $Z$ pole. 
$\effstw$ in the calculation is fixed to the 
average of the LEP-SLC measurement~\cite{LEP-SLC}. 
Results from the hadron colliders are excluded in order to avoid the influence from 
the PDFs used in those measurements. It would not have impact anyway because 
the hadron collider measurements~\cite{stw1, stw2, stw3, stw4} 
give $\effstw$ values very close to the LEP-SLC combination. 
The fixed $\Delta_u$ and $\Delta_d$ which describe the dependences on 
$Y$, $M$, and $Q_T$ in the intervals corresponding to the CMS measurement 
are predicted from the CT18 NNLO PDF~\cite{CT18}. 
We apply the same method to the previous CMS 8 TeV $\afb$ observation~\cite{CMSAFB8TeV}. 
However, compared to the previous work in Ref.~\cite{CMS8TeVExt}, 
we have fixed the $\effstw$ value in this study, as explained earlier. 
Our analysis confirms that varying the $\effstw$ value within its 
current uncertainty does not have noticeable effect on the determination of the 
proton parameters. 
By doing so, the uncertainties on the extracted $P_u$ and $P_d$ 
are reduced by roughly $50\%$. 
Due to its lower collision energy, the 8 TeV measurement corresponds 
to a larger $x$ value compared to the 13 TeV measurement.

\begin{figure}[!hbt]
\begin{center}
\epsfig{scale=0.4, file=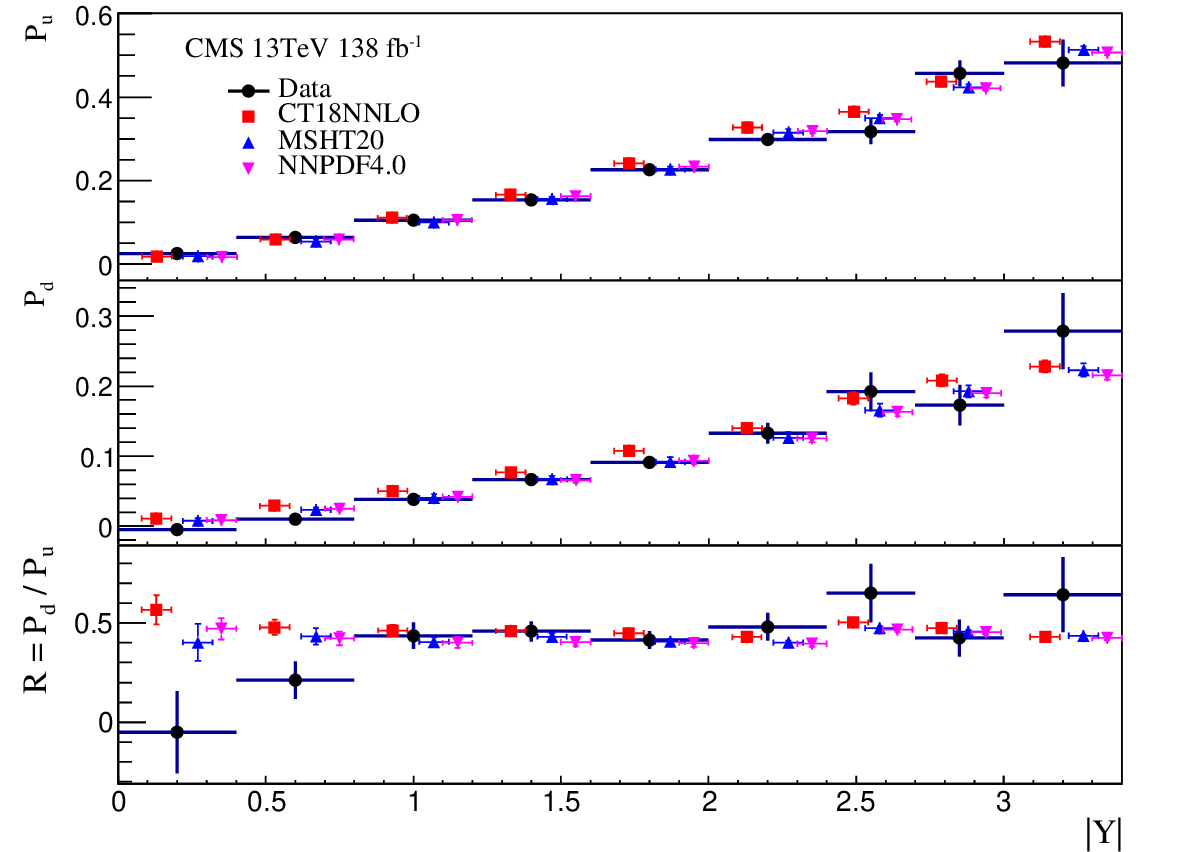}
\caption{\small $P_u$, $P_d$ and $R$ measured from the CMS 13 TeV $\afb$ in each 
$|Y|$ interval, compared to the predictions from CT18, MSHT20 and NNPDF4. 
Error bars of the measured $P_u$, $P_d$ and $R$ parameters correspond to the total uncertainties 
including those from experimental uncertainty extrapolation and the fixed $\Delta_u$ and $\Delta_d$. } 
\label{fig:PuPd13TeV}
\end{center}
\end{figure}

\begin{figure}[!hbt]
\begin{center}
\epsfig{scale=0.4, file=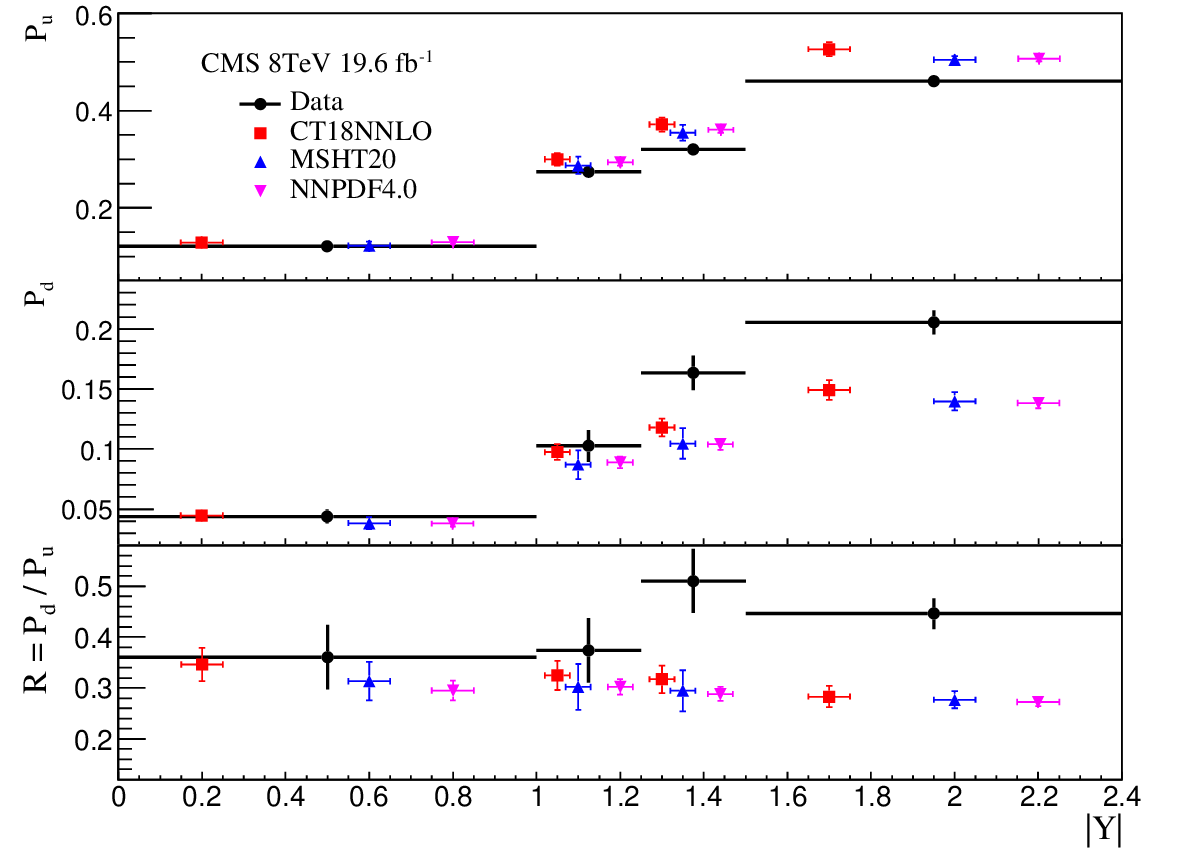}
\caption{\small $P_u$, $P_d$ and $R$ measured from the CMS 8 TeV $\afb$ in each 
$|Y|$ interval, compared to the predictions from CT18, MSHT20 and NNPDF4. 
Error bars of the measured $P_u$, $P_d$ and $R$ parameters correspond to the total uncertainties 
including those from experimental uncertainty extrapolation and the fixed $\Delta_u$ and $\Delta_d$. } 
\label{fig:PuPd8TeV}
\end{center}
\end{figure}

The extracted central values of $P_u$ and $P_d$ in each $|Y|$ interval, together with the predictions from 
various PDFs, are given in Fig.~\ref{fig:PuPd13TeV} for the 
13 TeV measurement and in Fig.~\ref{fig:PuPd8TeV} for the 8 TeV update, respectively.
Numerical results are listed in Table~\ref{tab:PuPd13TeV} and Table~\ref{tab:PuPd8TeV}. 
The uncertainties on the measured $P_u$ and $P_d$ consist of 3 parts: i) extrapolated from 
the experimental uncertainties on the original measured $\afb$, which is acquired directly 
in the minimal $\chi^2$ fitting and dominates the 
total uncertainty; ii) due to the fixed $\effstw$, which is estimated by varying its value 
according to the experimental uncertainty of the LEP-SLC determination and found to be 
negligible; iii) arising from the fixed 
$\Delta_u$ and $\Delta_d$, which can be computed using the PDF error sets provided by 
CT18. It has been checked that other PDFs, such as the MSHT20 and NNPDF4.0~\cite{MSHT20, NNPDF4}, 
give consistent predictions on $\Delta_u$ and $\Delta_d$ within the uncertainties~\cite{CMS8TeVExt, D0PuPd}. 
According to the CMS report, the experimental uncertainties on the observed $\afb$ are 
dominated by the statistical fluctuation. 
The uncertainties in the extracted $P_u$ and $P_d$ are highly 
correlated, as both parameters are simultaneously fitted from $A_\text{FB}$. 
These correlations should be accounted for when calculating 
the uncertainty in the $R$ parameter. Ideally, the correlation should 
reflect only the contributions from experimental uncertainties. However, 
using it as the total uncertainty correlation has no significant 
impact, as the $\Delta$-induced uncertainties remain small.


\begin{table}[hbt]
\begin{footnotesize}
\begin{center}
\begin{tabular}{l|c|c|c}
\hline \hline
 $|Y|$ & $P_u\pm(\text{exp.})\pm(\Delta)$ & $P_d\pm(\text{exp.})\pm(\Delta)$  & $\rho$ \\
\hline
 0-0.4 & $0.0242 \pm 0.0039\pm0.0004$ & $-0.0012 \pm 0.0052\pm0.0002$ & -0.88 \\
\hline
 0.4-0.8 & $0.0651 \pm0.0041\pm0.0005$ & $0.0138 \pm 0.0054\pm0.0006$ & -0.88  \\
\hline
 0.8-1.2 & $0.1054 \pm 0.0041\pm0.0004$ & $0.0460 \pm 0.0055\pm0.0008$ & -0.88  \\
\hline
 1.2-1.6 & $0.1574 \pm 0.0049\pm0.0006$ & $0.0720 \pm 0.0060\pm0.0007$  & -0.88 \\
\hline
 1.6-2.0 & $0.2307 \pm 0.0066\pm0.0010$ & $0.0956 \pm 0.0079\pm0.0011$  & -0.88 \\
\hline
 2.0-2.4 & $0.2966 \pm 0.0130\pm0.0012$ & $0.1423 \pm 0.0148\pm0.0021$  & -0.88 \\
\hline
 2.4-2.7 & $0.3159 \pm 0.0297\pm0.0018$ & $0.2054 \pm 0.0278\pm0.0038$  & -0.91 \\
\hline
 2.7-3.0 & $0.4503 \pm 0.0315\pm0.0026$ & $0.1907 \pm 0.0292\pm0.0043$  & -0.91\\
\hline
 3.0-3.4 & $0.4656 \pm 0.0557\pm0.0022$ & $0.2993 \pm 0.0542\pm0.0042$  & -0.92 \\
\hline \hline
\end{tabular}
\caption{\small Fitted values and uncertainties of $P_u$ and $P_d$ from the CMS 13 TeV $\afb$ measurement. 
The first uncertainties in the breakdown are extrapolated from the experimental uncertainties.  
The second uncertainties in the breakdown correspond to the 
theoretical errors arising from $\Delta_u$ and $\Delta_d$ estimated by using the CT18 error sets. 
The last column gives the correlation factor between the $P_u$ and $P_d$ uncertainties.}
\label{tab:PuPd13TeV}
\end{center}
\end{footnotesize}
\end{table}

\begin{table}[hbt]
\begin{footnotesize}
\begin{center}
\begin{tabular}{l|c|c|c}
\hline \hline
 $|Y|$ & $P_u\pm(\text{exp.})\pm(\Delta)$ & $P_d\pm(\text{exp.})\pm(\Delta)$ & $\rho$ \\
\hline
 0-1.0 & $0.1216 \pm 0.0062\pm0.0021$ & $0.0439 \pm 0.0071\pm0.0013$ & -0.88 \\
\hline
 1.0-1.25 & $0.2743 \pm0.0136\pm0.0033$ & $0.1025 \pm 0.0157\pm0.0025$ & -0.88  \\
\hline
 1.25-1.5 & $0.3205 \pm 0.0147\pm0.0033$ & $0.1636 \pm 0.0174 \pm0.0027$ & -0.88  \\
\hline
 1.5-2.4 & $0.4609 \pm 0.0132 \pm0.0021$ & $0.2056 \pm 0.0144\pm0.0030$ & -0.88  \\
\hline \hline
\end{tabular}
\caption{\small Fitted values and uncertainties of $P_u$ and $P_d$ from the CMS 8 TeV $\afb$ measurement. 
The first uncertainties in the breakdown are extrapolated from the experimental uncertainties. 
The second uncertainties in the breakdown correspond to the 
theoretical errors arising from $\Delta_u$ and $\Delta_d$ estimated by using the CT18 error sets. 
The last column gives the correlation factor between the $P_u$ and $P_d$ uncertainties.}
\label{tab:PuPd8TeV}
\end{center}
\end{footnotesize}
\end{table}

As shown in Fig.~\ref{fig:PuPd8TeV}, the measured $R$ values are significantly larger than current PDF predictions 
in the two high $|Y|$ intervals in the 8 TeV extraction, while in the two lower $|Y|$ intervals 
the measured values are consistent with predictions. According to Eq.~\eqref{eq:LOx}, 
the two high $|Y|$ intervals 
correspond to 
$x_1$ between $0.05$ and $0.1$, and $x_2$ around 0.002. For the two lower $|Y|$ intervals, 
$x_1\approx 0.02$ and $x_2\approx 0.005$. 
As demonstrated previously, the $R$ parameter is a 
direct observable on $d_V(x_1)/u_V(x_1)$, which 
can be sizably biased  
due to lack of directly experimental observation in current PDF
global analysis. 
The deviation is consistent with the D0 measurement~\cite{D0PuPd}, which gives $R$ 
at $x\sim 0.1$ larger than PDF predictions by 3.5 standard deviation. 

The 13 TeV extraction includes both overlapping and extended $x_1$ 
regions compared to the 8 TeV LHC and the D0 studies. In the 13 TeV data, 
the corresponding $x_2$ is even smaller, making the light quark distribution 
more consistent. 
For the high $|Y|$ intervals which correspond to 
$x_1>0.05$, $R$ is also measured to be in general larger than the predictions. 
However, the statistical fluctuation limits the precision. The most important results 
of the 13 TeV analysis come from the lower $|Y|$ intervals with high precision. 
In the $|Y|$ intervals between 
1.0 and 2.0, the $R$ parameter is precisely determined to be consistent with 
the predictions. It corresponds to $0.02<x<0.05$, 
the same region as the 8 TeV extraction, which also yields results consistent 
with the predictions.
The low $|Y|$ bins correspond to $x<0.02$, which is not covered by the 8 TeV 
kinematics. 
Within these intervals, the measured $R$ value is lower than the PDF 
prediction, reversing the trend described above. 
Note that the measured $P_d$ value is close to zero in the lowest $|Y|$ bin. 
According to Eq.~\eqref{eq:LHCAFBfactorization2}, it means $d$ and $\bar{d}$ 
quarks have almost consistent parton densities in the corresponding $x$ region. 
In another word, the measured structure parameter indicates that the 
perturbative QCD contribution is more significant with respect to the 
current PDF predictions.

According to Eq.~\eqref{eq:LHCAFBfactorization2} and the arguments following 
Eq.~\eqref{eq:LOx}, the impact of this 
new determination is clear that  
$R$ is mainly the ratio between $d_V(x_1)$ and $u_V(x_1)$. Focusing on the 
region where $x_1>0.05$, the data suggests a higher value 
of the down-to-up valence quark ratio, which can also be explained as 
either the $d/u$ ratio must increase or the $\bar{d}/\bar{u}$ SU(2) 
flavor asymmetry must be significantly suppressed. Similarly, 
at $x\sim 0.01$, an opposite shift in parton distributions is required 
to align with the observed data behavior. In other words, we conclude 
that the forward-backward asymmetries observed at the Tevatron and 
the LHC indicate a different $x$-dependence of the light-quark PDFs, 
particularly in the ratios of $d/u$ and $\bar{d}/\bar{u}$. These asymmetries 
can be sensitively probed only through this class of experimental 
observables, rather than through total cross-section measurements. 
In future global analyses of PDFs, incorporating the forward-backward 
asymmetry of Drell-Yan data will be crucial for enhancing our 
understanding of parton densities inside the proton. This is not 
only important for investigating light-quark PDFs but also for heavy-quark 
PDFs, 
through the interplay of various QCD sum rules, such as the 
momentum sum rule. 

\section{Acknowledgements}
This work was supported by the National Natural Science Foundation of China under Grant No. 11721505, 11875245, 
12061141005 and 12105275.  
C.-P. Yuan was supported by the U.S. National Science Foundation under Grant 
No. PHY-2310291.

\end{document}